\documentclass[letterpaper,english]{article}
\usepackage[T1]{fontenc}
\usepackage[latin1]{inputenc}
\usepackage{amsmath}
\usepackage{amssymb}

\makeatletter

\usepackage{graphicx}
\usepackage{amssymb}

\usepackage{amssymb}

\usepackage{amssymb}

\usepackage{graphicx}
\usepackage{geometry}

\usepackage{babel}

\usepackage{babel}

\usepackage{babel}

\usepackage{babel}
\makeatother
\begin{document}

\title{\textbf{The simple non-degenerate relativistic gas: statistical properties
and Brownian motion}}

\author{A. Sandoval-Villalbazo$^{1}$, A. Aragonés-Muñoz$^{1}$, \\
 A. L. García-Perciante
$^{2}$ \\$^{1}$Departamento de Física y Matemáticas\\ Universidad Iberoamericana \\
Prolongación Paseo de la Reforma 880, México D. F. 01210, México. \\
$^{2}$Depto. de Matemáticas Aplicadas y Sistemas \\ Universidad Autónoma
Metropolitana-Cuajimalpa \\ Artificios No. 40, México D. F. 01120, México.}

\maketitle

\begin{abstract}
 This paper shows a novel calculation of the mean square displacement of a classical Brownian particle in a relativistic thermal bath. The result is compared with the expressions obtained by other authors. Also, the thermodynamic properties of a non-degenerate simple relativistic gas are reviewed in terms of a treatment performed in velocity space.
\end{abstract}

\section{Introduction}

Interest in relativistic statistical mechanics has been revived due
to its applicability in several fields of study and the need of its internal theoretical consistence \cite{Cercignani}\cite{Groot}. New experimental devices, such as relativistic heavy ion colliders, have also drawn attention to this topic since they provide a new framework to test different approaches
to relativistic transport theory \cite{Ollitrault}. In the context of stochastic processes theory,  formalisms involving relativistic thermodynamics have been proposed in order to understand the nature of how relativity may broaden the usual conceptions of mathematical probability theory.

In this sense, the analysis of processes such as Brownian motion in a relativistic framework deserves a deeper look. Other issues related to the inclusion of non-equilibrium effects in relativity theory have been addressed in previous work \cite{San1}\cite{San2}. In the case of large (relativistic) temperatures, a description involving a Juttner function is required in order to compute the mean square displacement of Brownian particles \cite{Hänggi}. This calculation is described in the present work. The interest of Brownian motion has been recently revived due
to its applicability in different areas, including astrophysics \cite{Bertschinger}.

This paper is divided as follows. In section two, the basic properties
of a non-degenerate relativistic gas are reviewed and compared with
their Maxwell-Boltzmann function counterparts at mildly and ultra-relativistic
regimes. Section three is devoted to the study of relativistic Brownian
motion modeling the dynamics of a Newtonian brownian particle in a
relativistic thermal bath. Due to this aproximation, only elementary
mathematical tools are involved in the calculations. Finally, some theoretical
remarks on the use of relativistic statistical mechanics are included in section four.

\section{Distribution functions}

Some thermodynamic systems cannot be modeled
using the usual Maxwell-Boltzmann distribution. Instead, a more appropriate equilibrium
distribution function should be considered in high temperature cases. For example, the Jüttner distribution function for a special relativistic non-degenerate simple gas at rest is \cite{Cercignani}:
 \begin{equation}
f\left(\gamma;z\right)=\frac{e^{-\frac{\gamma}{z}}}{4\pi zc^{3}K_{2}\left(\frac{1}{z}\right)},\label{juttner}
\end{equation}
 where \begin{equation}
\gamma=\frac{1}{\sqrt{1-\frac{w^{2}}{c^{2}}}},\label{gamma}\end{equation}
 is the usual Lorentz factor for molecular speed $\omega$, $z=\frac{kT}{mc^{2}}$ is the relativistic
parameter at temperature $T$ for the system consisting of particles
of mass $m$, $K_{2}$ is the modified Bessel function of the second
kind, $k$ and $c$ are Boltzmann's constant and the speed of light,
respectively.

Figure 1 shows a comparison between both speed distribution functions as functions of $\beta=\frac{\omega}{c}$,
corresponding to Eq. (\ref{juttner}) and a Maxwell-Boltzmann distribution written in terms of $\gamma$:

\begin{equation}
f_{MB}\left(\gamma;z\right)=\left(\frac{1}{2\pi zc^2}\right)^{3/2}\frac{4\pi c^2(\gamma^{2}-1)}{\gamma^{2}}e^{-\frac{{\gamma^{2}}-1}{2z\gamma^{2}}}\label{MB}=n \left(\frac{m}{2\pi kT}\right)^{\frac{3}{2}} e^{-\frac{mv^2}{2kT}},
\end{equation}
 for an electron gas. Although the differences seem small, they are measurable and can be detected in certain astrophysical processes such as the Sunyaev-Zel'dovich effect \cite{Rephaeli}.

\begin{figure}

\caption{This figure shows the probability distribution functions, for $z=0.01$
(left) and $z=0.1$ (right), both in the classical case (dotted line) and
in the relativistic case (solid line). The small differences at $z=0.01$,
corresponding to an electron temperature of $6\times10^{7}K$ for an electron
gas, are large enough to account for relativistic corrections to the Sunyaev-Zel'dovich
effect in clusters of galaxies. The relativistic features of the probability function are significantly enhanced
for $z=0.1$, corresponding to an electron temperature of $6\times10^{8}K$.}

\begin{centering}
\includegraphics[
  width=7in,
  height=2.5in]{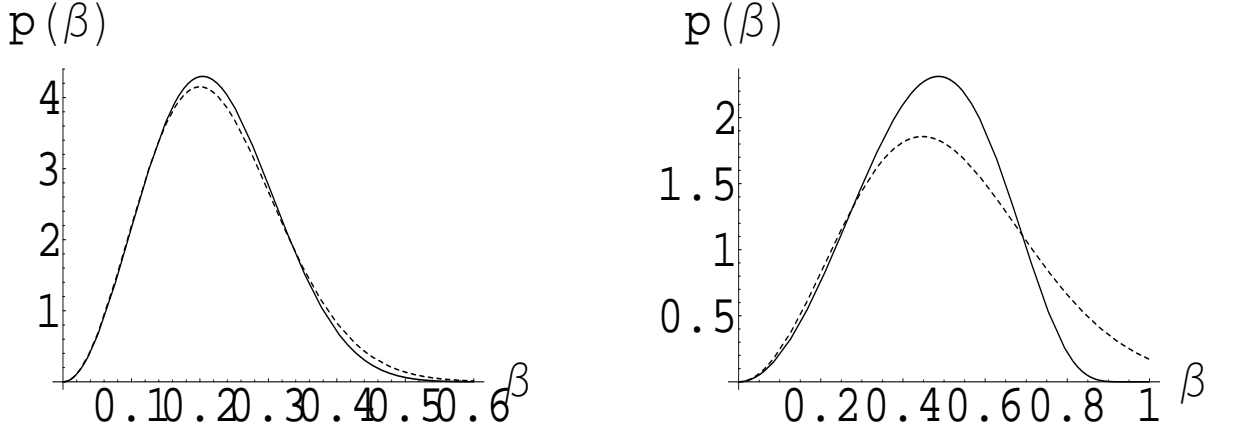}

\end{centering}

\label{fig:F1}
\end{figure}

 In special relativity, the four-vector $v^{\nu}$ is related to the three velocity
$\vec{w}$ by:

\begin{equation}
v^{\nu}=\left(\begin{array}{c}
\gamma\vec{w}\\
\gamma c\end{array}\right),\label{vel3}\end{equation}
 and the volume element in velocity space reads $d^{3}v=4\pi w^{2}\gamma^{5}dw$
\cite{Liboff}. One fundamental difference between both distributions,
Maxwellian and Jüttnerian, is that the former allows for molecular
speeds greater than $c$, while the latter has a cutoff at $\beta=1$. Both behaviors are shown in Fig. 1. On the first plot both distributions have a similar qualitative behavior,
where most particles lie on the right side of the peaks. However,
for larger values of $z$, as can be seen in the second plot, both
distributions are quite different. Notice how the dotted,
classical, distribution exceeds the physical limit $\beta=1$ while
the Jüttner distribution function vanishes exactly at that value.

We are now in position to address a simple application of relativistic
thermodynamics and the corresponding comparison with the non-relativistic
case.

\section{A simple approach to relativistic Brownian motion}

Before addressing the calculation of the mean internal
energy per particle for a non-degenerate gas, both in the non-relativistic
and relativistic cases. In the non-relativistic case, this quantity
is identified with the average integral:

\begin{equation}
<\epsilon>=\int_{0}^{\infty}\frac{1}{2}mv^{2}f_{MB}(v)dv=\frac{1}{2}m^{\frac{5}{2}}\left(\frac{1}{2\pi z}\right)^{3/2}\int_{0}^{\infty}v^{4}e^{\frac{-mv^{2}}{2kT}}dv.\label{eninternanr}\end{equation}
 The usual change of variable $\omega=\sqrt{\frac{m}{2kT}}v$ leads
to the result:

\begin{equation}
<\epsilon>=\frac{4}{\sqrt{\pi}}kT\int_{0}^{\infty}\omega^{4}e^{-\omega^{2}}d\omega=\frac{3kT}{2}.
\label{eninternanr2}
\end{equation}
 This result is identified with the energy equipartition theorem.
Each degree of freedom contributes $\frac{kT}{2}$ units of energy
to the total internal energy per particle of the system. As we shall
see below, the energy equipartition theorem does not hold in relativistic
thermodynamics.

In the relativistic case, since the kinetic energy per particle is
$mc^{2}\gamma$, the counterpart of Eq.(\ref{eninternanr}) reads:

\begin{equation}
<\epsilon>=\int_{0}^{c}mc^{2}\gamma f(w)(4\pi w^{2}\gamma^{5})dw.\label{enintr}\end{equation}
 In terms of $\gamma$ Eq. (\ref{enintr}) reads:

\begin{equation}
<\epsilon>=\frac{mc^{2}}{zK_{2}\left(\frac{1}{z}\right)}\int_{1}^{\infty}\gamma^{2}e^{\frac{-\gamma}{z}}(\gamma^{2}-1)^{\frac{1}{2}}d\gamma,\label{enintr2}\end{equation}
 which yields an expression for $<\epsilon>$ in terms of modified
Bessel functions of the second kind, namely \cite{Cercignani}
\begin{equation}
<\epsilon>=3kT+mc^{2}\frac{K_{1}\left(\frac{1}{z}\right)}{K_{2}\left(\frac{1}{z}\right)}.
\label{enint3}
\end{equation}

Equation (\ref{enint3}) reflects the fact that, for low enough temperatures,
internal energy is well described by a linear function of the temperature
$T$, and a non-linear behavior is exhibited for increasing $z$. The
use of asymptotic expansions for the modified Bessel functions present
in this equation, for $z\ll1$ leads to the result of Eq. (\ref{eninternanr2}) \cite{Abramowitz}.

In order to account for the mean square displacement $<r^{2}>$ of
a \emph{classical} Brownian particle of mass $M$ in a relativistic
gas, we assume that the relativistic parameter $z$ of the gas is
not negligible, while the equation of motion describing the displacement
still reads:

\begin{equation}
M\frac{d\vec{v}}{dt}+\mu\vec{v}=\vec{F}_{a},
\label{mov1}
\end{equation}
 where $F_{a}$ is a random force, $\mu$ is the dynamic friction
coefficient and $\overrightarrow{v}$ is the velocity of the particle. As usual, we assume that $<\vec{F}_{a}>=\vec{0}$ so that
  \begin{equation}
\frac{\mu}{2}\frac{d}{dt}\langle r^{2}\rangle=\langle\epsilon\rangle,
\label{prommu}
\end{equation}

In Eq. (\ref{prommu}) we will now use the expression given by (\ref{enint3}), for $<\epsilon>$, since the relativistic parameter is significant. We finally get,
following the usual steps to compute the mean-square displacement
\cite{Feynman}:
\begin{equation}
\langle[r(t)]^{2}\rangle=\frac{2}{\mu}\left[3kT+mc^{2}\frac{K_{1}\left(\frac{1}{z}\right)}{K_{2}\left(\frac{1}{z}\right)}\right]t,
\label{promr}
\end{equation}
 which reduces to the well-known Einstein-Smoluchowsky equation
for the non-relativistic case, if we substract the rest energy to the internal energy given by Eq. (\ref{enint3}). Historically, this equation has been
used to establish the value of Boltzmann's constant $k$ from direct
measurement of the mean square displacement in various experimental
situations \cite{Bertschinger}. In figure 2, we plot the next equation in order to compare the relativistic mean square displacement with its non-relativistic counterpar:
\begin{equation}
\langle\frac{r^2}{2Dt}\rangle=2(1+\frac{1}{3z}\frac{K_1(\frac{1}{z})}{K_2(\frac{1}{z})}-\frac{1}{3z}),
\label{desvplot}
\end{equation}
where $D=\frac{3kT}{\mu}$. For $z\ll1$ we obtain the correct non-relativistic limit
\begin{equation}
\langle\frac{r^2}{2Dt}\rangle\sim1.
\label{limit}
\end{equation}

\begin{figure}

\caption{Comparison of the values of the mean square displacement for relativistic
and non-relativistic temperatures. For different values of $z$ the asymptotic
behavior of the ratio tends to $1$.}

\begin{centering}
\includegraphics[
  width=7in,
  height=4in]{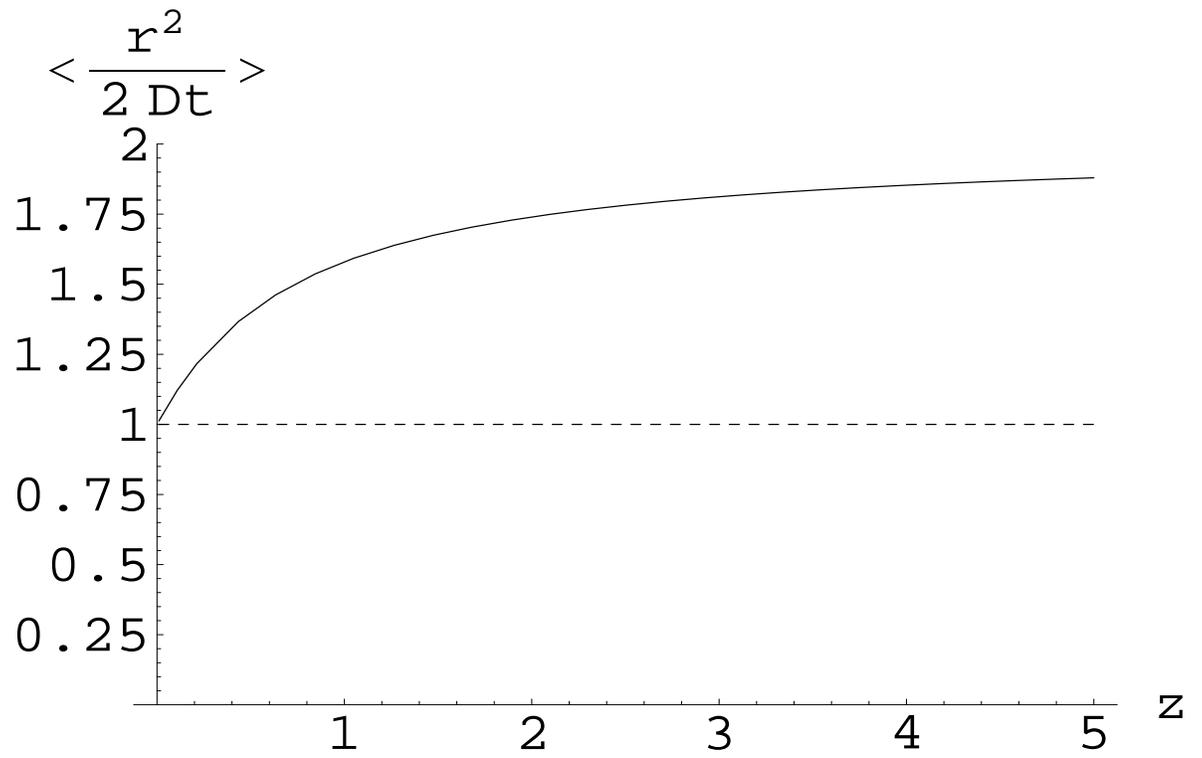}

\end{centering}

\label{fig:F3}
\end{figure}

\section{Final remarks}

The framework here presented provides a direct prediction for
the value of the mean square displacement of a massive particle exhibiting
Brownian motion in a mildly relativistic, simple gas. There is an
elementary physical argument to support the approach here proposed.
Brownian motion is caused by random collisions of light particles
with massive particles. In a certain range of temperatures, the
relativistic parameter $Z=\frac{kT}{Mc²}$ for the massive particles
can be considered negligible, allowing a classical description for
their dynamics, while for the thermal bath, $z\simeq1$
is significant. Since a non-relativistic distribution predicts an
important fraction of particles with supraluminal velocities for large
$z$, a relativistic treatment describing the properties of the gas
is certainly needed.  Other calculations of the mean square displacement in this physical situation
may require the use of numerical methods or more elaborated approaches \cite{Hänggi2}.

A similar calculation has been permormed in reference \cite{Hänggi2} in the case of a Wiener process in which $r^2$ is time-dependent for a given value of $z$. Work in this direction will be permormed in the future.

\end{document}